\documentclass[letterpaper,twocolumn,english,prb]{revtex4}
\usepackage[T1]{fontenc}
\usepackage[latin9]{inputenc}
\usepackage{color}
\usepackage{babel}
\usepackage{array}
\usepackage{amsmath}
\usepackage{amssymb}
\usepackage{graphicx}
\usepackage{esint}
\usepackage[unicode=true,pdfusetitle,
 bookmarks=true,bookmarksnumbered=false,bookmarksopen=false,
 breaklinks=false,pdfborder={0 0 1},backref=false,colorlinks=false]
 {hyperref}

\makeatletter

\pdfpageheight\paperheight
\pdfpagewidth\paperwidth

\providecommand{\tabularnewline}{\\}

\@ifundefined{textcolor}{}
{%
 \definecolor{BLACK}{gray}{0}
 \definecolor{WHITE}{gray}{1}
 \definecolor{RED}{rgb}{1,0,0}
 \definecolor{GREEN}{rgb}{0,1,0}
 \definecolor{BLUE}{rgb}{0,0,1}
 \definecolor{CYAN}{cmyk}{1,0,0,0}
 \definecolor{MAGENTA}{cmyk}{0,1,0,0}
 \definecolor{YELLOW}{cmyk}{0,0,1,0}
}

\usepackage{babel}
\usepackage{babel}
\usepackage{babel}
\usepackage{babel}
\usepackage{babel}
\usepackage{babel}
\usepackage{babel}

\newcommand{\nimpc}{n_{\rm imp,c}}
\newcommand{\vsc}{V_{\rm D,sc}}

\makeatother

\begin{document}

\title{Disorder-tuned selection of order in bilayer graphene}

\author{Junhua Zhang,$^{1}$ Rahul Nandkishore,$^{2,3}$ and E. Rossi$^{1}$}

\affiliation{$^{1}$Department of Physics, College of William and Mary, Williamsburg,
Virginia 23187, USA\\
 $^{2}$Department of Physics, Massachusetts Institute of Technology,
Cambridge, Massachusetts 02139, USA\\
 $^{3}$Princeton Center for Theoretical Science, Princeton University,
Princeton, New Jersey 08544, USA}

\date{\today}
\begin{abstract}
The nature of the interaction-driven spontaneously broken-symmetry
state in charge neutral bilayer graphene (BLG) has attracted a lot
of interest. Theoretical studies predict various ordered states as
the candidates for the ground state of BLG in the absence of external
fields. Several experiments have been performed by different groups
to identify the nature of the collective ground state in BLG. However,
so far, there is no consensus: some experiments show evidence that
suggests the establishment of a nematic gapless state, while others
present results that are more consistent with the establishment of
a fully gapped state. Moreover, even among the experiments that appear
to see a bulk gap, some of the samples are found to be conducting
(suggesting existence of gapless edge states), while others are insulating.
Here we explore the hypothesis that disorder might explain the discrepancy
between experiments. We find that the pair-breaking effect due to
non-magnetic short-range disorder varies among the candidate ground
states, giving rise to different amounts of suppression of their mean-field
transition temperatures. Our results indicate that BLG can undergo
a transition between different ordered states as a function of the
disorder strength providing a possible scenario to resolve the discrepancy
between experimental observations. 
\end{abstract}
\maketitle

\section{introduction}

AB-stacked bilayer graphene (BLG) \cite{novoselov2006,mccann2012,neto2009,dassarma2011}
is formed by two graphene \cite{novoselov2004} layers rotated by
60$^{o}$ with respect to each other. Its low-energy band structure
is characterized by parabolic conduction and valence bands that touch
at the corners, the $K$ and $K'$ points, of the Brillouin zone.
A number of theoretical works have predicted various spontaneously-broken-symmetry
states as the candidates for the ground state of BLG near the charge
neutrality point (CNP) in the absence of external fields. \cite{min2008b,fzhang2010,nandkishore2010,nandkishore2010c,vafek2010,lemonik2010,vafek2010b,fzhang2011,fzhang2012,nandkishore2012,lemonik2012,gorbar2012,lang2012}
The multiple degrees of freedom in BLG -- layer, spin, and valley
-- give rise to the diversity of the candidate orders. In general,
the proposed ordered states can be classified in two groups: (i) gapped
states characterized by the opening of a full gap in the quasiparticle
spectrum, and (ii) nematic states in which the quadratic band crossing
points at which the conduction and valence bands touch are split into
two Dirac points leaving the quasiparticle spectrum gapless. These
two groups have a different structure with respect to the layer index:
gapped states are layer-polarized while nematic states are not.\cite{nandkishore2012}
Depending on the valley and spin structure different collective states
can be identified in each general group. Gapped states with different
spin-valley structures include the quantum valley Hall (QVH), the
quantum anomalous Hall (QAH) and the quantum spin Hall (QSH) state,
as well as a layer antiferromagnet (LAF) state. Within mean field
theory, in the clean limit, the states in each group have the same
transition temperature, $T_{c,0}^{G}$ for the gapped states, and
$T_{c,0}^{N}$ for the nematic states.

Several experimental groups have made efforts to ascertain the nature
of the ground state using high-quality suspended BLG.\cite{martin2010,weitz2010,mayorov2011,freitag2011,velasco2011,bao2012,veligura2012,freitag2013}
They all find evidence of spontaneous symmetry breaking at low temperatures.
However, they reach different conclusions on the identity of the
ordered state: First, some experiments show evidence that supports
the establishment of a nematic state,\cite{mayorov2011} while others
either present results that are more consistent with the establishment
of a gapped state \cite{freitag2011,velasco2011,bao2012,veligura2012,freitag2013}
or are consistent with both type of states \cite{martin2010,weitz2010};
Second, among the experiments supporting the establishment of a gapped
state, some indicate that the gapped state comes with conducting
edge states \cite{martin2010,weitz2010,freitag2011,freitag2013} and
others indicate that the state is fully insulating \cite{freitag2011,velasco2011,bao2012,veligura2012,freitag2013}
e.g. the LAF state. One explanation that has been proposed for this
multitude of conflicting experimental results is that BLG is highly
multi critical,\cite{cvetkovic2012} and that different experimental
samples fall in the basin of attraction of different correlated fixed
points. 

One important and unavoidable factor present in all materials that
has the potential to strongly affect the formation and nature of a
broken symmetry state is disorder, due, for instance, to charge impurities,
adatoms, vacancies, and ripples. For example, it is well known that
the presence of magnetic impurities in BCS superconductors can strongly
decrease the transition temperature ($T_{c}$).\cite{abrikosov1961,maki1969}
The pair-breaking effect of magnetic impurities in BCS superconductors
can be attributed to the different scattering off the impurities of
the time-reversed fermionic states forming the Cooper pairs. Another
example is the pair-breaking effect of normal impurities on exciton
condensates.\cite{zittartz1967,bistritzer2008a} Since the broken-symmetry
states in BLG involve particle-hole pairing with different layer-spin-valley
structures, we expect that different pairing structures could be affected
differently by disorder. 

\begin{figure}[t]
\hfill{}\includegraphics[width=8.5cm]{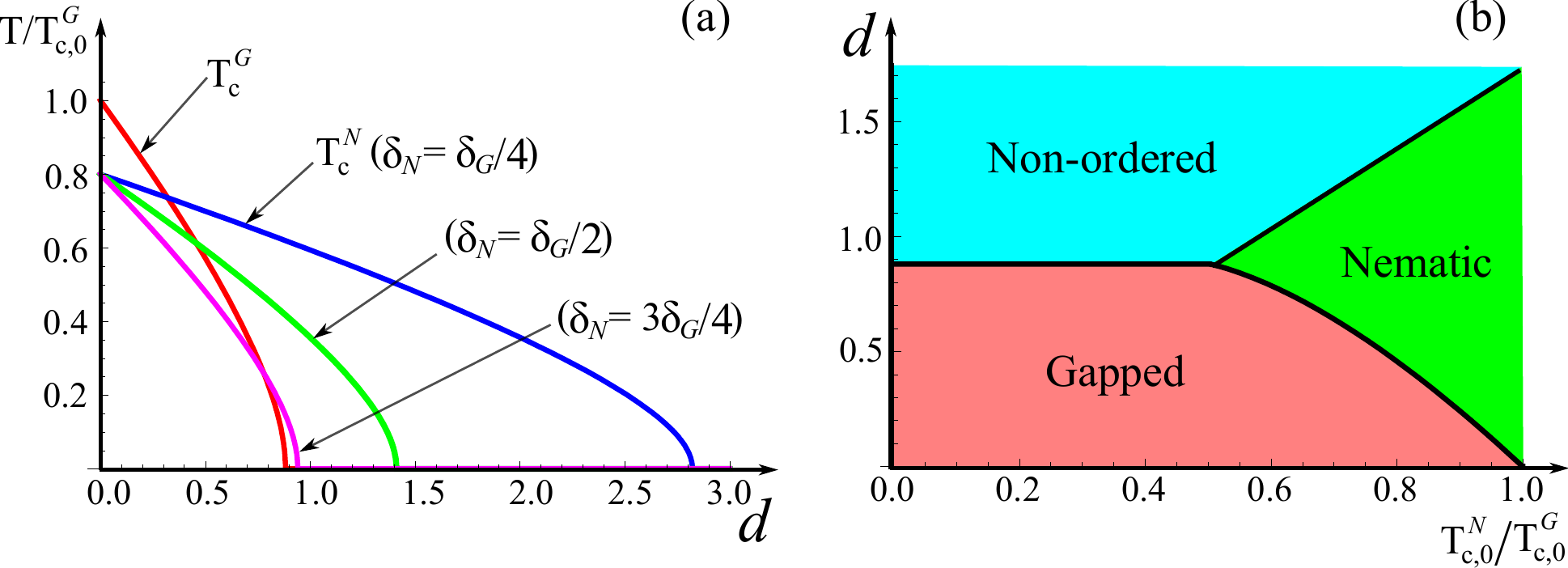}\hfill{}

\caption{(Color online) (a) The mean-field phase transition temperatures, $T_{c}^{G}$
for the gapped phase and $T_{c}^{N}$ for the nematic phase under
three interlayer disorder correlation conditions, are plotted as functions
of the intra valley disorder strength by solving Eq.\,(\ref{eq: universal_relation})
for the case that the clean-limit transition temperatures of the two
phases have the relation: $T_{c,0}^{N}/T_{c,0}^{G}=0.8$. (b) Phase
diagram obtained by calculating the critical disorder strength for
various ratios of $T_{c,0}^{N}/T_{c,0}^{G}$ for the case of uncorrelated
disorder in which $\delta_{N}=\delta_{G}/2$.}

\label{fig: transition} 
\end{figure}

In this work, we study the effect of disorder on the broken-symmetry
states of BLG near the CNP in the absence of external fields. We consider
only non-magnetic disorder and do not take into account spin flip
scattering. Within mean field theory, in the clean limit, the transition
temperature of the gapped phase is higher than that of the nematic
phase. However, we find that this scenario can be modified when the
presence of disorder is taken into account. Considering non-magnetic
short-range disorder, we find that in the presence of disorder that
causes intra valley scattering only, the transition temperature of
the gapped states is suppressed more than the transition temperature
of the nematic states. Thus, within mean field theory, our results
indicate that below a critical strength of disorder the system is
prone to be in a gapped phase whereas above the critical disorder
strength the nematic phase is favored, as shown in Fig.\,\ref{fig: transition}.
In addition, we find that non-magnetic disorder producing inter valley
scattering also contributes to the suppression of $T_{c}$ for the
valley-unpolarized gapped states but does not affect $T_{c}$ for
the valley-polarized gapped states. Since valley-polarized gapped
states have co-propagating edge modes in the two valleys (which cannot
be gapped out in the absence of magnetic disorder), while valley-unpolarized
gapped states have counter propagating edge modes (which can be gapped
out in the presence of inter valley scattering), our results on the
effect of inter valley disorder could also be part of the explanation
of why some experiments see conducting states with a bulk gap while
others see insulating gapped states.

\section{theory and results}

At low energies, the mean-field Hamiltonian ($\hat{H}$) that describes
a broken-symmetry state of BLG can be written as: $\hat{H}=\hat{H}_{0}+\hat{\Delta}+\hat{V}$
where 
\begin{equation}
\hat{H}_{0}(\mathbf{k})\!=\!\left[\begin{array}{cc}
\hat{h}(\mathbf{k}) & 0\\
0 & \hat{h}^{*}(-\mathbf{k})
\end{array}\right]\!;\,\hat{h}(\mathbf{k})\!=\!\left[\begin{array}{cc}
-\mu & \varepsilon_{k}e^{-i2\theta_{\mathbf{k}}}\\
\varepsilon_{k}e^{i2\theta_{\mathbf{k}}} & -\mu
\end{array}\right],\label{eq: model}
\end{equation}
$\hat{V}$ is the non-magnetic disorder potential, $\mathbf{k}=(k_{x},k_{y})$,
$\theta_{\mathbf{k}}=\arctan(k_{y}/k_{x})$, and $\varepsilon_{k}\equiv\frac{\hbar^{2}k^{2}}{2m^{*}}$
with $m^{*}\approx0.03m_{e}$. $\hat{H}_{0}$ is degenerate in spin
space, and $\hat{h}$ is a $2\times2$ matrix in layer space. Current
experiments reveal that the relevant energy scale for the broken symmetry
state is of the order of few meV,\cite{martin2010,weitz2010,mayorov2011,freitag2011,velasco2011,bao2012,veligura2012,freitag2013}
and that in the absence of magnetic field the instability toward an
ordered state is the strongest at the CNP. \cite{min2008b,fzhang2010,nandkishore2010,nandkishore2010c,vafek2010,lemonik2010,vafek2010b,fzhang2011,fzhang2012,nandkishore2012,lemonik2012,gorbar2012,lang2012}
As a consequence, for our purposes the low-energy two-band model (\ref{eq: model})
is adequate and in addition we can focus our attention to the case
when the chemical potential $\mu$ is fixed at the CNP (i.e., $\mu=0$).
The two groups of candidate ordered states are distinguished by the
structure in layer space of the order parameter: $\hat{\Delta}=\Delta_{G}\hat{\sigma}_{z}$
for the gapped states and $\hat{\Delta}=\Delta_{N}\hat{\sigma}_{x}$
for the nematic states (without loss of generality we have chosen
the complex nematic order parameter $\Delta_{N}$ to be real), where
$\hat{\sigma}$'s are Pauli matrices acting on the layer space. Taking
into account the valley degree of freedom, we have $\hat{\Delta}=\Delta_{G}\hat{\sigma}_{z}\hat{\tau}_{0}$
($\Delta_{N}\hat{\sigma}_{x}\hat{\tau}_{0}$) for the gapped (nematic)
valley-independent states, and $\hat{\Delta}=\Delta_{G}\hat{\sigma}_{z}\hat{\tau}_{z}$
($\Delta_{N}\hat{\sigma}_{x}\hat{\tau}_{z}$) for the gapped (nematic)
valley-polarized states, where $\hat{\tau}$'s are Pauli matrices
acting on the valley space. The disorder potential can be written
in the general form $\hat{V}=\hat{U}+\hat{W},$ with $\hat{U}\sim U_{\sigma}\delta_{\sigma\sigma'}\hat{\tau}_{0}$
and $\hat{W}\sim W_{\sigma}\delta_{\sigma\sigma'}(\hat{\tau}_{x}+i\hat{\tau}_{y})/2+h.c.$,
where $U_{\sigma}$ ($W_{\sigma}$, $W_{\sigma}^{*}$) is the intra
(inter) valley disorder potential in layer $\sigma$. 

The influence of disorder is taken into account using the self-consistent
Born approximation. After averaging over disorder realizations, the
effect of disorder is captured by the self-energy matrix $\hat{\Sigma}$
that renormalizes the quasiparticle propagator and the pairing vertex
of the condensate.

\subsection{Intra valley disorder scattering}

We first consider the case in which disorder-induced valley-flip scattering
processes are negligible, i.e., $\hat{W}=0$. In this case, our discussion
can be simplified to the $2\times2$ layer space since intra valley
scattering does not lift the degeneracy between ground states that
differ in valley structure. The renormalized Green's function $\hat{\mathcal{G}}$
is given by 
\begin{equation}
\hat{\mathcal{G}}(\mathbf{k},i\omega_{n})=\left[i\omega_{n}\hat{\sigma}_{0}-\hat{h}(\mathbf{k})-\hat{\Delta}-\hat{\Sigma}(\mathbf{k},i\omega_{n})\right]^{-1},
\end{equation}
where $\omega_{n}=(2n+1)\pi T$ are the Matsubara frequencies, $T$
is the temperature, and 
\begin{equation}
\Sigma_{\sigma\sigma'}(\mathbf{k},i\omega_{n})=n_{U}\int\frac{d^{2}\mathbf{p}}{(2\pi)^{2}}U_{\sigma,\mathbf{k-p}}\mathcal{G}_{\sigma\sigma'}(\mathbf{p},i\omega_{n})U_{\sigma',\mathbf{p-k}}
\end{equation}
is the disorder-averaged self-energy. Here $n_{U}$ is the density
of the randomly-distributed intra valley scattering centers. It is
reasonable to assume $n_{U}$ to be the same in the two layers.

For the gapped states, the self-consistency equation for the order
parameter takes the form 
\begin{equation}
\Delta_{G}=-\frac{1}{2}\Gamma_{S}T\sum_{n}\int\frac{d^{2}\mathbf{k}}{(2\pi)^{2}}\mathrm{Tr}\left[\hat{\sigma}_{z}\hat{\mathcal{G}}(\mathbf{k},i\omega_{n})\right],\label{eq: gap_equation_gapped}
\end{equation}
where $\Gamma_{S}$ is the effective coupling and $\mathrm{Tr}[\dots]$
is the trace of the argument. The disorder renormalized Green's function
can be written as 
\begin{equation}
\hat{\mathcal{G}}_{G}(\mathbf{k},i\omega_{n})=\left[\begin{array}{cc}
i\tilde{\omega}_{n}-\tilde{\Delta}_{G} & -\varepsilon_{k}e^{-i2\theta_{\mathbf{k}}}\\
-\varepsilon_{k}e^{i2\theta_{\mathbf{k}}} & i\tilde{\omega}_{n}+\tilde{\Delta}_{G}
\end{array}\right]^{-1},
\end{equation}
where 
\begin{align}
\tilde{\omega}_{n} & =\omega_{n}+n_{U}\int\frac{d^{2}\mathbf{p}}{(2\pi)^{2}}\left|U_{\mathbf{k-p}}\right|^{2}\frac{\tilde{\omega}_{n}}{\tilde{\omega}_{n}^{2}+\varepsilon_{p}^{2}+\tilde{\Delta}_{G}^{2}},\nonumber \\
\tilde{\Delta}_{G} & =\Delta_{G}-n_{U}\int\frac{d^{2}\mathbf{p}}{(2\pi)^{2}}\left|U_{\mathbf{k-p}}\right|^{2}\frac{\tilde{\Delta}_{G}}{\tilde{\omega}_{n}^{2}+\varepsilon_{p}^{2}+\tilde{\Delta}_{G}^{2}}.\label{eq: renormalization_gapped}
\end{align}
In the above expressions we have assumed that the disorder strength
is the same in the two layers, i.e., $|U_{\mathbf{k-p}}|\equiv\left|U_{1,\mathbf{k-p}}\right|=\left|U_{2,\mathbf{k-p}}\right|$.
In the case of short-range disorder potential, $U_{\sigma,\mathbf{k-p}}=U$,
we obtain 
\begin{eqnarray}
\tilde{\omega}_{n} & = & \omega_{n}+\frac{1}{2}\left(\frac{1}{\tau_{2}}+\frac{1}{\tau_{1}}\right)\frac{\tilde{\omega}_{n}}{\sqrt{\tilde{\omega}_{n}^{2}+\tilde{\Delta}_{G}^{2}}},\nonumber \\
\tilde{\Delta}_{G} & = & \Delta_{G}-\frac{1}{2}\left(\frac{1}{\tau_{2}}-\frac{1}{\tau_{1}}\right)\frac{\tilde{\Delta}_{G}}{\sqrt{\tilde{\omega}_{n}^{2}+\tilde{\Delta}_{G}^{2}}},\label{eq: renormalize_gapped}
\end{eqnarray}
where $\frac{1}{\tau_{1}}$ and $\frac{1}{\tau_{2}}$ are the collision
rates resulting from the disorder potential. In this case, $\frac{1}{\tau_{2}}=n_{U}U^{2}\frac{m^{*}}{2\hbar^{2}}$
and $\frac{1}{\tau_{1}}=0$. Note that the opposite sign in front
of $\frac{1}{\tau_2}$ in the equations for $\tilde{\omega}_{n}$
and $\tilde{\Delta}_{G}$ gives rise to the pair-breaking effect of
disorder on the condensate. On the other hand, the term proportional
to $\frac{1}{\tau_{1}}$ in Eq.\,(\ref{eq: renormalize_gapped})
has the same sign in the equations for $\tilde{\omega}_{n}$ and $\tilde{\Delta}_{G}$,
and consequently $\frac{1}{\tau_{1}}$ does not affect the transition temperature. We
can therefore see that for the gapped state the effect of intra valley
disorder is analogous to the effect of magnetic impurities on BCS
superconductors,\cite{abrikosov1961} which is purely pair-breaking.
From Eq.\,(\ref{eq: gap_equation_gapped}) and (\ref{eq: renormalize_gapped})
the mean-field critical temperature $T_{c}$ in the presence of disorder
is given by a universal function in terms of the pair-breaking parameter
$\delta=1/\tau_{2}$ \cite{abrikosov1961}, 
\begin{equation}
\ln\left[\frac{T_{c,0}}{T_{c}}\right]=\psi\left(\frac{1}{2}+\frac{\delta}{2\pi T_{c}}\right)-\psi\left(\frac{1}{2}\right),\label{eq: universal_relation}
\end{equation}
where $\psi(z)$ is the di-gamma function, and $T_{c,0}$ is the transition
temperature in the clean limit. For the gapped phase $T_{c,0}=T_{c,0}^{G}$
is given by
\begin{equation}
k_{B}T_{c,0}^{G}=\frac{2}{\pi}\gamma E_{c}\exp\left[-\frac{4\pi\hbar^{2}}{\Gamma_{S}m^{*}}\right],\label{eq: Tc0_gapped}
\end{equation}
where $\gamma\approx1.78$ is the Euler's constant, and $E_{c}$ is
a cutoff for the energy range of the interaction. The value of the
pair-breaking parameter $\delta$ is $\delta_{G}=\frac{1}{\tau_{2}}=n_{U}U^{2}\frac{m^{*}}{2\hbar^{2}}$
for the gapped states. When $\delta_{G}/2\pi T_{c}\ll1$, the transition
temperature is linearly suppressed: $T_{c}^{G}=T_{c,0}^{G}-\frac{\pi}{4}\delta_{G}$.
The critical disorder strength, above which the gapped phase is completely
suppressed, is given by $\delta_{c}^{G}=\pi/(2\gamma)T_{c,0}^{G}\approx0.88T_{c,0}^{G}$.
Assuming that the dominant source of disorder is charge impurities,
\cite{dassarma2011} using the condition $\delta_{c}^{G}=0.88T_{c,0}^{G}$,
we can provide a quantitative estimate of the critical value of the
impurity density $\nimpc$ above which $T_{c}\to0$. Taking into account
screening effects the effective, screened, disorder potential $\vsc$
due to the charge impurities is short-range with strength $U(q)=\vsc(q)=2\pi e^{2}/(\kappa q\epsilon(q))$,
where $\kappa$ is the dielectric constant and $\epsilon(q)$ is the
dielectric function. For $q<2k_{F}$ we have \cite{dassarma2011}
$U=\vsc(q<2k_{F})=2\pi\hbar^{2}/(g_{s}g_{v}m^{*})$ where $g_{s}=g_{v}=2$
are the spin and valley degeneracy respectively. We then find (set
$k_{B}\equiv1$ here) the critical impurity density: 
\begin{equation}
n_{\text{imp,c}}^{G}=\frac{4}{\gamma\pi}\frac{m^{*}}{\hbar^{2}}T_{c,0}^{G}=3\times10^{10}{\rm cm}^{-2}\frac{T_{c,0}^{G}}{{\rm meV}}.\label{eq:nimpc}
\end{equation}
Experimentally for the gapped phase $T_{c,0}^{G}$ appears to be on
the order of 1~meV.\cite{bao2012} Eq.~(\ref{eq:nimpc}) then allows
us to predict that in order to have the establishment of the gapped
phase the impurity density has to be lower than $\sim3\times10^{10}{\rm cm}^{-2}$.
This estimate is consistent with current experiments, see in particular
Ref.\ {[}\onlinecite{martin2010,weitz2010}{]}. In addition, Eq.~(\ref{eq:nimpc})
allows to obtain $T_{c,0}^{G}$, a quantity that is very difficult
to estimate accurately, by knowing the value of $n_{\text{imp,c}}^{G}$.

For the nematic states, the self-consistent equation for the order
parameter takes the form 
\begin{equation}
\Delta_{N}=-\frac{1}{2}\Gamma_{D}T\sum_{n}\int\frac{d^{2}\mathbf{k}}{(2\pi)^{2}}\mathrm{Tr}\left[\hat{\sigma}_{x}\hat{\mathcal{G}}(\mathbf{k},i\omega_{n})\right],\label{eq:gap_equation_nematic}
\end{equation}
where $\Gamma_{D}$ is the effective interlayer coupling. The renormalized
Green's function after averaging over disorder can be written as 
\begin{equation}
\hat{\mathcal{G}}_{N}(\mathbf{k},i\omega_{n})=\left[\begin{array}{cc}
i\tilde{\omega}_{n} & -\varepsilon_{k}e^{-i2\theta_{\mathbf{k}}}-\tilde{\Delta}_{N}\\
-\varepsilon_{k}e^{i2\theta_{\mathbf{k}}}-\tilde{\Delta}_{N} & i\tilde{\omega}_{n}
\end{array}\right]^{-1},
\end{equation}
where 
\begin{align}
 & \tilde{\omega}_{n}=\omega_{n}+n_{U}\times\nonumber \\
 & \int\frac{d^{2}\mathbf{p}}{(2\pi)^{2}}\left|U_{\mathbf{k-p}}\right|^{2}\frac{\tilde{\omega}_{n}}{\tilde{\omega}_{n}^{2}+\varepsilon_{p}^{2}+\tilde{\Delta}_{N}^{2}+2\varepsilon_{p}\tilde{\Delta}_{N}\cos(2\theta_{\mathbf{p}})},\nonumber \\
 & \tilde{\Delta}_{N}=\Delta_{N}-n_{U}\times\nonumber \\
 & \int\frac{d^{2}\mathbf{p}}{(2\pi)^{2}}U_{1,\mathbf{k-p}}U_{2,\mathbf{k-p}}^{*}\frac{\varepsilon_{p}e^{-i2\theta_{\mathbf{p}}}+\tilde{\Delta}_{N}}{\tilde{\omega}_{n}^{2}+\varepsilon_{p}^{2}+\tilde{\Delta}_{N}^{2}+2\varepsilon_{p}\tilde{\Delta}_{N}\cos(2\theta_{\mathbf{p}})}.\label{eq: renormalization_nematic}
\end{align}
Here again we assumed $\left|U_{1}\right|=\left|U_{2}\right|$. In
order to discuss the influence of disorder on $T_{c}$ we evaluate
Eq.\,(\ref{eq: renormalization_nematic}) in the limit $T\rightarrow T_{c}$,
where the order parameter becomes vanishingly small, $\Delta_{N}\rightarrow0$.
Assuming short-range disorder, $U_{\sigma,\mathbf{k-p}}=U_{\sigma}$,
to leading order in $\Delta_{N}$ we obtain (for $\tilde{\omega}_{n}>0$),
\begin{eqnarray}
\tilde{\omega}_{n} & = & \omega_{n}+\frac{1}{2}\left(\frac{1}{\tau_{2}}+\frac{1}{\tau_{1}}\right)\frac{\tilde{\omega}_{n}}{\tilde{\omega}_{n}},\nonumber \\
\tilde{\Delta}_{N} & = & \Delta_{N}-\frac{1}{2}\left(\frac{1}{\tau_{2}}-\frac{1}{\tau_{1}}\right)\frac{\tilde{\Delta}_{N}}{\tilde{\omega}_{n}}.\label{eq: renormalize_nematic}
\end{eqnarray}
Linearizing Eq.\,(\ref{eq:gap_equation_nematic}) near $T_{c}$,
we again find that the transition temperature satisfies Eq. (\ref{eq: universal_relation}),
with the pair-breaking parameter $\delta_{N}=\frac{1}{\tau_{2}}$.
In the limit $\delta_{N}/2\pi T_{c}\ll1$, the transition temperature
is linearly suppressed: $T_{c}^{N}=T_{c,0}^{N}-\frac{\pi}{4}\delta_{N}$.
The critical disorder strength, above which the nematic phase is completely
destroyed, is given by $\delta_{c}^{N}\approx0.88T_{c,0}^{N}$. Notice
that both the clean limit transition temperature and the value of
the pair-breaking parameter are different from the ones obtained for
the gapped phase. For the nematic phase, the mean-field transition
temperature in the clean limit is given by 
\begin{equation}
k_{B}T_{c,0}^{N}=\frac{2}{\pi}\gamma E_{c}\exp\left[-\frac{8\pi\hbar^{2}}{\Gamma_{D}m^{*}}\right].\label{eq: Tc0_nematic}
\end{equation}
Notice that assuming $\Gamma_{D}\approx\Gamma_{S}$, Eq.~(\ref{eq: Tc0_nematic})
and (\ref{eq: Tc0_gapped}) imply $T_{c,0}^{N}<T_{c,0}^{G}$. Equation
(\ref{eq: renormalization_nematic}) shows that the renormalized quantity
$\tilde{\Delta}_{N}$ depends on the correlation property between
the disorder potentials in the two layers: (i) When the disorder potentials
in the two layers are perfectly correlated: $U_{1}=U_{2}\equiv U$,
we have $\frac{1}{\tau_{2}}=n_{U}U^{2}\frac{3m^{*}}{8\hbar^{2}},\ \frac{1}{\tau_{1}}=n_{U}U^{2}\frac{m^{*}}{8\hbar^{2}}$,
so that $\delta_{N}=n_{U}U^{2}\frac{3m^{*}}{8\hbar^{2}}$. In this
case the relation between the pair-breaking parameter values in the
two phases is $\delta_{N}=\frac{3}{4}\delta_{G}$. (ii) When the disorder
potentials of the two layers are uncorrelated: $\Sigma_{12}=\Sigma_{21}=0$,
then $\frac{1}{\tau_{2}}-\frac{1}{\tau_{1}}=0$. In the limit $T\rightarrow T_{c}$,
$\frac{1}{\tau_{2}}=\frac{1}{\tau_{1}}=n_{U}U^{2}\frac{m^{*}}{4\hbar^{2}}$,
and we find $\delta_{N}=n_{U}U^{2}\frac{m^{*}}{4\hbar^{2}}$. In this
case we have the relation $\delta_{N}=\frac{1}{2}\delta_{G}$. (iii)
When the disorder potentials in the two layers are perfectly anticorrelated:
$U_{1}=-U_{2}$, in the limit $T\rightarrow T_{c}$, we have $\frac{1}{\tau_{2}}=n_{U}U^{2}\frac{m^{*}}{8\hbar^{2}},\ \frac{1}{\tau_{1}}=n_{U}U^{2}\frac{3m^{*}}{8\hbar^{2}}$.
In this case we find $\delta_{N}=n_{U}U^{2}\frac{m^{*}}{8\hbar^{2}}$,
so that $\delta_{N}=\frac{1}{4}\delta_{G}$. 

\begin{table}[t]
\begin{tabular}{|c|c|c|c|}
\hline 
$\delta/\delta_{G}$  & correlated  & uncorrelated  & anticorrelated\tabularnewline
\hline 
\hline 
Gapped phase  & 1  & 1  & 1\tabularnewline
\hline 
Nematic phase  & 3/4  & 1/2  & 1/4\tabularnewline
\hline 
\end{tabular}\caption{Comparison of the magnitudes of pair-breaking effect in the gapped
and the nematic phase under different interlayer disorder correlation
conditions. \label{tab: table1}}
\end{table}

We summarize the magnitudes of the pair-breaking effect of disorder in the gapped
and in the nematic phase under different interlayer disorder correlation
conditions in Table \ref{tab: table1}. Irrespective of the interlayer
correlations of disorder, the disorder suppression of $T_{c}$ is
weaker in the nematic phase than in the gapped phase. Assuming $T_{c,0}^{N}<T_{c,0}^{G}$,
we then find that the system can undergo a transition from the gapped
phase to the nematic gapless phase by changing the strength of disorder,
as shown in Fig.\,\ref{fig: transition}. Figure \ref{fig: transition}
(a) shows $T_{c}$, obtained by solving Eq.\,(\ref{eq: universal_relation}),
as a function of the intra valley disorder strength characterized
by the dimensionless variable $d\equiv n_{U}U^{2}m^{*}/2\hbar^{2}T_{c,0}^{G}$,
for the case of $T_{c,0}^{N}/T_{c,0}^{G}=0.8$, in the gapped and
the nematic phase under the three interlayer disorder correlation
conditions. Below a critical disorder strength the gapped phase is
dominant while above it the nematic phase becomes preferable. The
phase diagram calculated at various $d$ and $T_{c,0}^{N}/T_{c,0}^{G}$
in the case of $\delta_{N}=\delta_{G}/2$ is shown in Fig.\,\ref{fig: transition}
(b). 

If the dominant source of disorder is charge impurities, analogous
to Eq.~(\ref{eq:nimpc}) we can then provide a quantitative estimate
for the critical impurity density $n_{\text{imp,c}}^{N}$, above which
the nematic phase is completely suppressed. We find 
\begin{equation}
n_{\text{imp,c}}^{N}=A\times3\times10^{10}{\rm cm}^{-2}\frac{T_{c,0}^{N}}{{\rm meV}},\label{eq: nematic_nimpc}
\end{equation}
where $A=4/3$, 2, or 4 depending on the interlayer correlation properties of disorder.

\subsection{Inter valley disorder scattering}

\begin{table}[t]
\begin{tabular}{|c|>{\centering}p{3.4cm}|>{\centering}p{3.8cm}|}
\hline 
 & valley-polarized states (QAH, QSH)  & valley-independent states (LAF, QVH)\tabularnewline
\hline 
\hline 
$\delta/\delta_{G}$  & 1  & $1+\frac{n_{W}\left|W\right|^{2}}{n_{U}U^{2}}$\tabularnewline
\hline 
\end{tabular}\caption{Comparison of the magnitudes of pair-breaking effect between different
valley-structured varieties of the gapped states. \label{tab: table2} }
\end{table}

In this section, we discuss the effect of inter valley disorder, i.e.,
$\hat{W}\neq0$. In this case the resulting valley-flip processes distinguish between states
with different valley structure. In the following we consider the
case in which the two types of disorder potential $\hat{U}$ and $\hat{W}$
are uncorrelated, and $\left|U_{1}\right|=\left|U_{2}\right|\equiv U$,
$\left|W_{1}\right|=\left|W_{2}\right|\equiv\left|W\right|$, and
the density of inter valley scattering centers $n_{W}$ is the same
in the two layers.

In the gapped phase, taking into account the presence of inter valley
scattering, for the valley-independent states (LAF, QVH) the scattering
rates in Eq. (\ref{eq: renormalize_gapped}) become: $\frac{1}{\tau_{2}}=\left(n_{U}U^{2}+n_{W}\left|W\right|^{2}\right)\frac{m^{*}}{2\hbar^{2}},\ \frac{1}{\tau_{1}}=0$,
indicating an enhancement on the pair-breaking effect characterized
by $\delta_{G,v}=\left(n_{U}U^{2}+n_{W}\left|W\right|^{2}\right)\frac{m^{*}}{2\hbar^{2}}=\delta_{G}\left(1+\frac{n_{W}\left|W\right|^{2}}{n_{U}U^{2}}\right)$.
On the other hand, for the valley-polarized states (QAH, QSH), we
obtain $\frac{1}{\tau_{2}}=n_{U}U^{2}\frac{m^{*}}{2\hbar^{2}},\ \frac{1}{\tau_{1}}=n_{W}\left|W\right|^{2}\frac{m^{*}}{2\hbar^{2}}$,
indicating that the pair-breaking effect is unaltered since the influence
of the inter valley disorder only introduces a non pair-breaking component
$\frac{1}{\tau_{1}}$.

Table \ref{tab: table2} summarizes the effect of inter valley disorder
on the different gapped states. Our results suggest that the valley-independent
states (LAF, QVH) are more likely to appear in samples with very low
disorder while the valley-polarized states (QAH, QSH) could survive
at higher disorder concentrations. 

For the nematic phase we find that if $W_{1}$ and $W_{2}$ are uncorrelated,
states with different valley structure are equally affected and therefore
the inter valley disorder does not favor a specific valley-structure.

\section{connection to current experiments}

Currently, two experimental groups have conducted
comparative studies on samples with different disorder strengths:
(i) The measurements presented in Ref.~{[}\onlinecite{freitag2011}{]},
performed on suspended and current annealed BLG devices, reveal two
kinds of samples, B1 and B2. B2 samples are found to be gapped with
vanishingly small conductance at the CNP in zero external fields,
while B1 samples exhibit a small but finite conductance. The measurements
show that B2 samples are \textit{cleaner}
than B1 samples; (ii) The most systematic study is done in Ref.~{[}\onlinecite{bao2012}{]}.
In this work the authors investigate twenty-three high-quality suspended
BLG devices and find that these samples, at low temperatures ($T<10$
K) and zero external fields, fall into two groups: sixteen samples
have a minimum conductivity of the order of $2-3\ e^{2}/h$, whereas
seven samples are practically insulating with conductivity $\le0.4\ e^{2}/h$.
At the same time, the seven insulating samples are among the highest
room-temperature mobility samples, indicating a lower disorder strength
in the insulating samples. Notice that the value of the minimum conductivity
($2-3\ e^{2}/h$) reported in Ref. {[}\onlinecite{bao2012}{]} for the sixteen
samples with lower mobility ($3\times10^{4}-10^{5}\ \mathrm{cm}^{2}/\mathrm{Vs}$)
is quite smaller than the value of minimum conductivity expected for
samples of this quality in the normal (non ordered) state of BLG.\cite{dassarma2011}
It is then natural to expect that these sixteen samples, at low temperature,
might be in a nematic or a gapped valley-polarized state and not in
the normal state.

It is a possiblel scenario to interpret the results presented in Ref.\ {[}\onlinecite{freitag2011, bao2012}{]} as suggesting that the cleanest
samples are in a valley-independent gapped state that has no protected
edge currents (insulating) and that the samples with lower mobility, higher disorder
strength, are either in the nematic gapless phase or in a gapped valley-polarized
state that has protected edge currents. This interpretation of the
measurements of these comparative experimental studies is qualitatively
consistent with our results that show that as the strength of the
non-magnetic disorder increases the valley-independent gapped states
get suppressed more strongly and the nematic or the gapped valley-polarized
states become favored. In addition, in the experiments presented in
Ref.\ {[}\onlinecite{martin2010,weitz2010}{]} it is estimated that the
density of charge impurities in the sample that exhibits signatures
of a broken symmetry phase, is on the order of $10^{10}{\rm cm}^{-2}$.
This order of impurity density is consistent with our results given
that it is lower than the value that we obtain, Eq.\,(\ref{eq:nimpc},
\ref{eq: nematic_nimpc}), for the critical charge impurity density, above
which $T_{c}\rightarrow0$, for both the gapped and the nematic phase,
considering that in the clean limit $T_{c}$ is on the order of few
meV.

The discussion above indicates that the effect of disorder described
in our work should be directly relevant to current experiments on
BLG, with some limitations. The experimental results presented in
Ref.\ {[}\onlinecite{freitag2011, bao2012}{]}
clearly show that disorder plays an important role in determining
the nature of the broken-symmetry state in BLG. Our work provides
an insight on how non-magnetic disorder might resolve the competition
between different ordered states. Given the difficulty of probing
experimentally the nature of the ordered phase, the strength of the
disorder, and in particular the relative strength of inter valley
and intra valley disorder, more work is needed to fully characterize
the effect of the interplay between electron correlations and disorder
in BLG.

\section{conclusions}

In conclusion, we have studied the effect of non-magnetic disorder
on the nature of bilayer graphene broken symmetry state that is expected
to be established when the chemical potential is set at the charge
neutrality point even in the absence of external electric and magnetic
fields. Current experiments have shown signatures suggesting that
the broken symmetry state could be either in a gapped
phase or in a nematic gapless phase. For this reason we focused
our analysis only on these two groups of ordered states, even though
it has been shown theoretically that many other competing ordered
states are possible. \cite{min2008b,fzhang2010,nandkishore2010,nandkishore2010c,vafek2010,lemonik2010,vafek2010b,fzhang2011,fzhang2012,nandkishore2012,lemonik2012,gorbar2012,lang2012}

We find that in the presence of intra valley disorder, the resulting
pair-breaking effects have different magnitude in the gapped and in
the nematic phase: the transition temperature is suppressed more strongly
in the gapped phase than in the nematic phase. Moreover, we find that
in the nematic phase the pair-breaking effect of the disorder depends
significantly on the interlayer correlation properties of the disorder:
the pair-breaking effect is weaker in the uncorrelated case than in
the perfectly correlated case, and it is the weakest for the case
of perfectly anticorrelated disorder. We also find that the presence
of inter valley disorder enhances the pair-breaking effect of disorder
on the valley-independent gapped states but that it merely contributes
a non pair-breaking component to the valley-polarized gapped states. 

Our results suggest that clean BLG might have a valley-independent
gapped ground state (e.g. LAF), which does not have protected edge
modes, but that small amounts of inter valley disorder can drive it
into a valley-polarized gapped state with edge modes (e.g. QAH or
QSH), and that intra valley disorder can drive it into a nematic state.
The relation of our results to the current available experiments has
been discussed. In addition, assuming charge impurities to be the
dominant source of disorder, we provide a quantitative estimate of
the critical impurity densities above which the gapped and the nematic
order vanish, which can be tested in experiments. 
\begin{acknowledgments}
We would like to thank Leonid Levitov for numerous helpful discussions;
RN would like to thank Leonid Levitov also for a long collaboration
on bilayer graphene. JZ and ER acknowledge support by ONR, Grant No.
ONR-N00014-13-1-0321, and the Jeffress Memorial Trust. RN acknowledges
support from a PCTS fellowship.
\end{acknowledgments}


\begin{thebibliography}{31}
\expandafter\ifx\csname natexlab\endcsname\relax\def\natexlab#1{#1}\fi
\expandafter\ifx\csname bibnamefont\endcsname\relax
  \def\bibnamefont#1{#1}\fi
\expandafter\ifx\csname bibfnamefont\endcsname\relax
  \def\bibfnamefont#1{#1}\fi
\expandafter\ifx\csname citenamefont\endcsname\relax
  \def\citenamefont#1{#1}\fi
\expandafter\ifx\csname url\endcsname\relax
  \def\url#1{\texttt{#1}}\fi
\expandafter\ifx\csname urlprefix\endcsname\relax\def\urlprefix{URL }\fi
\providecommand{\bibinfo}[2]{#2}
\providecommand{\eprint}[2][]{\url{#2}}

\bibitem[{\citenamefont{Novoselov et~al.}(2006)\citenamefont{Novoselov, McCann,
  Morozov, Falko, Katsnelson, Zeitler, Jiang, Schedin, and
  Geim}}]{novoselov2006}
\bibinfo{author}{\bibfnamefont{K.}~\bibnamefont{Novoselov}},
  \bibinfo{author}{\bibfnamefont{E.}~\bibnamefont{McCann}},
  \bibinfo{author}{\bibfnamefont{S.}~\bibnamefont{Morozov}},
  \bibinfo{author}{\bibfnamefont{V.}~\bibnamefont{Falko}},
  \bibinfo{author}{\bibfnamefont{M.}~\bibnamefont{Katsnelson}},
  \bibinfo{author}{\bibfnamefont{U.}~\bibnamefont{Zeitler}},
  \bibinfo{author}{\bibfnamefont{D.}~\bibnamefont{Jiang}},
  \bibinfo{author}{\bibfnamefont{F.}~\bibnamefont{Schedin}}, \bibnamefont{and}
  \bibinfo{author}{\bibfnamefont{A.}~\bibnamefont{Geim}},
  \bibinfo{journal}{Nature Physics} \textbf{\bibinfo{volume}{2}},
  \bibinfo{pages}{177} (\bibinfo{year}{2006}).

\bibitem[{\citenamefont{{McCann} and {Koshino}}(2013)}]{mccann2012}
\bibinfo{author}{\bibfnamefont{E.}~\bibnamefont{{McCann}}} \bibnamefont{and}
  \bibinfo{author}{\bibfnamefont{M.}~\bibnamefont{{Koshino}}},
  \bibinfo{journal}{Reports on Progress in Physics}
  \textbf{\bibinfo{volume}{76}}, \bibinfo{pages}{056503}
  (\bibinfo{year}{2013}).

\bibitem[{\citenamefont{Neto et~al.}(2009)\citenamefont{Neto, Guinea, Peres,
  Novoselov, and Geim}}]{neto2009}
\bibinfo{author}{\bibfnamefont{A.~H.~C.} \bibnamefont{Neto}},
  \bibinfo{author}{\bibfnamefont{F.}~\bibnamefont{Guinea}},
  \bibinfo{author}{\bibfnamefont{N.~M.~R.} \bibnamefont{Peres}},
  \bibinfo{author}{\bibfnamefont{K.~S.} \bibnamefont{Novoselov}},
  \bibnamefont{and} \bibinfo{author}{\bibfnamefont{A.~K.} \bibnamefont{Geim}},
  \bibinfo{journal}{Rev. Mod. Phys.} \textbf{\bibinfo{volume}{81}},
  \bibinfo{pages}{109} (\bibinfo{year}{2009}).

\bibitem[{\citenamefont{{Das Sarma} et~al.}(2011)\citenamefont{{Das Sarma},
  Adam, Hwang, and Rossi}}]{dassarma2011}
\bibinfo{author}{\bibfnamefont{S.}~\bibnamefont{{Das Sarma}}},
  \bibinfo{author}{\bibfnamefont{S.}~\bibnamefont{Adam}},
  \bibinfo{author}{\bibfnamefont{E.~H.} \bibnamefont{Hwang}}, \bibnamefont{and}
  \bibinfo{author}{\bibfnamefont{E.}~\bibnamefont{Rossi}},
  \bibinfo{journal}{Rev. Mod. Phys.} \textbf{\bibinfo{volume}{83}},
  \bibinfo{pages}{407} (\bibinfo{year}{2011}).

\bibitem[{\citenamefont{Novoselov et~al.}(2004)\citenamefont{Novoselov, Geim,
  Morozov, Jiang, Zhang, Dubonos, Grigorieva, and Firsov}}]{novoselov2004}
\bibinfo{author}{\bibfnamefont{K.~S.} \bibnamefont{Novoselov}},
  \bibinfo{author}{\bibfnamefont{A.~K.} \bibnamefont{Geim}},
  \bibinfo{author}{\bibfnamefont{S.~V.} \bibnamefont{Morozov}},
  \bibinfo{author}{\bibfnamefont{D.}~\bibnamefont{Jiang}},
  \bibinfo{author}{\bibfnamefont{Y.}~\bibnamefont{Zhang}},
  \bibinfo{author}{\bibfnamefont{S.~V.} \bibnamefont{Dubonos}},
  \bibinfo{author}{\bibfnamefont{I.~V.} \bibnamefont{Grigorieva}},
  \bibnamefont{and} \bibinfo{author}{\bibfnamefont{A.~A.}
  \bibnamefont{Firsov}}, \bibinfo{journal}{Science}
  \textbf{\bibinfo{volume}{306}}, \bibinfo{pages}{666} (\bibinfo{year}{2004}).

\bibitem[{\citenamefont{Min et~al.}(2008)\citenamefont{Min, Borghi, Polini, and
  MacDonald}}]{min2008b}
\bibinfo{author}{\bibfnamefont{H.}~\bibnamefont{Min}},
  \bibinfo{author}{\bibfnamefont{G.}~\bibnamefont{Borghi}},
  \bibinfo{author}{\bibfnamefont{M.}~\bibnamefont{Polini}}, \bibnamefont{and}
  \bibinfo{author}{\bibfnamefont{A.~H.} \bibnamefont{MacDonald}},
  \bibinfo{journal}{Phys. Rev. B} \textbf{\bibinfo{volume}{77}},
  \bibinfo{pages}{041407(R)} (\bibinfo{year}{2008}).

\bibitem[{\citenamefont{Zhang et~al.}(2010)\citenamefont{Zhang, Min, Polini,
  and MacDonald}}]{fzhang2010}
\bibinfo{author}{\bibfnamefont{F.}~\bibnamefont{Zhang}},
  \bibinfo{author}{\bibfnamefont{H.}~\bibnamefont{Min}},
  \bibinfo{author}{\bibfnamefont{M.}~\bibnamefont{Polini}}, \bibnamefont{and}
  \bibinfo{author}{\bibfnamefont{A.~H.} \bibnamefont{MacDonald}},
  \bibinfo{journal}{Phys. Rev. B} \textbf{\bibinfo{volume}{81}},
  \bibinfo{pages}{041402(R)} (\bibinfo{year}{2010}).

\bibitem[{\citenamefont{Nandkishore and
  Levitov}(2010{\natexlab{a}})}]{nandkishore2010}
\bibinfo{author}{\bibfnamefont{R.}~\bibnamefont{Nandkishore}} \bibnamefont{and}
  \bibinfo{author}{\bibfnamefont{L.}~\bibnamefont{Levitov}},
  \bibinfo{journal}{Phys. Rev. Lett.} \textbf{\bibinfo{volume}{104}},
  \bibinfo{pages}{156803} (\bibinfo{year}{2010}{\natexlab{a}}).

\bibitem[{\citenamefont{Nandkishore and
  Levitov}(2010{\natexlab{b}})}]{nandkishore2010c}
\bibinfo{author}{\bibfnamefont{R.}~\bibnamefont{Nandkishore}} \bibnamefont{and}
  \bibinfo{author}{\bibfnamefont{L.}~\bibnamefont{Levitov}},
  \bibinfo{journal}{Phys. Rev. B} \textbf{\bibinfo{volume}{82}},
  \bibinfo{pages}{115124} (\bibinfo{year}{2010}{\natexlab{b}}).

\bibitem[{\citenamefont{Vafek and Yang}(2010)}]{vafek2010}
\bibinfo{author}{\bibfnamefont{O.}~\bibnamefont{Vafek}} \bibnamefont{and}
  \bibinfo{author}{\bibfnamefont{K.}~\bibnamefont{Yang}},
  \bibinfo{journal}{Phys. Rev. B} \textbf{\bibinfo{volume}{81}},
  \bibinfo{pages}{041401(R)} (\bibinfo{year}{2010}).

\bibitem[{\citenamefont{Lemonik et~al.}(2010)\citenamefont{Lemonik, Aleiner,
  Toke, and Fal'ko}}]{lemonik2010}
\bibinfo{author}{\bibfnamefont{Y.}~\bibnamefont{Lemonik}},
  \bibinfo{author}{\bibfnamefont{I.~L.} \bibnamefont{Aleiner}},
  \bibinfo{author}{\bibfnamefont{C.}~\bibnamefont{Toke}}, \bibnamefont{and}
  \bibinfo{author}{\bibfnamefont{V.~I.} \bibnamefont{Fal'ko}},
  \bibinfo{journal}{Phys. Rev. B} \textbf{\bibinfo{volume}{82}},
  \bibinfo{pages}{201408(R)} (\bibinfo{year}{2010}).

\bibitem[{\citenamefont{Vafek}(2010)}]{vafek2010b}
\bibinfo{author}{\bibfnamefont{O.}~\bibnamefont{Vafek}},
  \bibinfo{journal}{Phys. Rev. B} \textbf{\bibinfo{volume}{82}},
  \bibinfo{pages}{205106} (\bibinfo{year}{2010}).

\bibitem[{\citenamefont{Zhang et~al.}(2011)\citenamefont{Zhang, Jung, Fiete,
  Niu, and H.}}]{fzhang2011}
\bibinfo{author}{\bibfnamefont{F.}~\bibnamefont{Zhang}},
  \bibinfo{author}{\bibfnamefont{J.}~\bibnamefont{Jung}},
  \bibinfo{author}{\bibfnamefont{G.~A.} \bibnamefont{Fiete}},
  \bibinfo{author}{\bibfnamefont{Q.}~\bibnamefont{Niu}}, \bibnamefont{and}
  \bibinfo{author}{\bibfnamefont{M.~A.} \bibnamefont{H.}},
  \bibinfo{journal}{Phys. Rev. Lett.} \textbf{\bibinfo{volume}{106}},
  \bibinfo{pages}{156801} (\bibinfo{year}{2011}).

\bibitem[{\citenamefont{Zhang et~al.}(2012)\citenamefont{Zhang, Min, and
  MacDonald}}]{fzhang2012}
\bibinfo{author}{\bibfnamefont{F.}~\bibnamefont{Zhang}},
  \bibinfo{author}{\bibfnamefont{H.}~\bibnamefont{Min}}, \bibnamefont{and}
  \bibinfo{author}{\bibfnamefont{A.~H.} \bibnamefont{MacDonald}},
  \bibinfo{journal}{Phys. Rev. B} \textbf{\bibinfo{volume}{86}},
  \bibinfo{pages}{155128} (\bibinfo{year}{2012}).

\bibitem[{\citenamefont{Nandkishore and Levitov}(2012)}]{nandkishore2012}
\bibinfo{author}{\bibfnamefont{R.}~\bibnamefont{Nandkishore}} \bibnamefont{and}
  \bibinfo{author}{\bibfnamefont{L.}~\bibnamefont{Levitov}},
  \bibinfo{journal}{Phys. Scr.} \textbf{\bibinfo{volume}{T146}},
  \bibinfo{pages}{014011} (\bibinfo{year}{2012}).

\bibitem[{\citenamefont{Lemonik et~al.}(2012)\citenamefont{Lemonik, Aleiner,
  and Fal'ko}}]{lemonik2012}
\bibinfo{author}{\bibfnamefont{Y.}~\bibnamefont{Lemonik}},
  \bibinfo{author}{\bibfnamefont{I.}~\bibnamefont{Aleiner}}, \bibnamefont{and}
  \bibinfo{author}{\bibfnamefont{V.~I.} \bibnamefont{Fal'ko}},
  \bibinfo{journal}{Phys. Rev. B} \textbf{\bibinfo{volume}{85}},
  \bibinfo{pages}{245451} (\bibinfo{year}{2012}).

\bibitem[{\citenamefont{Gorbar et~al.}(2012)\citenamefont{Gorbar, Gusynin,
  Miransky, and Shovkovy}}]{gorbar2012}
\bibinfo{author}{\bibfnamefont{E.~V.} \bibnamefont{Gorbar}},
  \bibinfo{author}{\bibfnamefont{V.~P.} \bibnamefont{Gusynin}},
  \bibinfo{author}{\bibfnamefont{V.~A.} \bibnamefont{Miransky}},
  \bibnamefont{and} \bibinfo{author}{\bibfnamefont{I.~A.}
  \bibnamefont{Shovkovy}}, \bibinfo{journal}{Phys. Rev. B}
  \textbf{\bibinfo{volume}{86}}, \bibinfo{pages}{125439}
  (\bibinfo{year}{2012}).

\bibitem[{\citenamefont{Lang et~al.}(2012)\citenamefont{Lang, Meng, Scherer,
  Uebelacker, Assaad, Muramatsu, Honerkamp, and Wessel}}]{lang2012}
\bibinfo{author}{\bibfnamefont{T.~C.} \bibnamefont{Lang}},
  \bibinfo{author}{\bibfnamefont{Z.~Y.} \bibnamefont{Meng}},
  \bibinfo{author}{\bibfnamefont{M.~M.} \bibnamefont{Scherer}},
  \bibinfo{author}{\bibfnamefont{S.}~\bibnamefont{Uebelacker}},
  \bibinfo{author}{\bibfnamefont{F.~F.} \bibnamefont{Assaad}},
  \bibinfo{author}{\bibfnamefont{A.}~\bibnamefont{Muramatsu}},
  \bibinfo{author}{\bibfnamefont{C.}~\bibnamefont{Honerkamp}},
  \bibnamefont{and} \bibinfo{author}{\bibfnamefont{S.}~\bibnamefont{Wessel}},
  \bibinfo{journal}{Phys. Rev. Lett.} \textbf{\bibinfo{volume}{109}},
  \bibinfo{pages}{126402} (\bibinfo{year}{2012}).

\bibitem[{\citenamefont{Martin et~al.}(2010)\citenamefont{Martin, Feldman,
  Weitz, Allen, and Yacoby}}]{martin2010}
\bibinfo{author}{\bibfnamefont{J.}~\bibnamefont{Martin}},
  \bibinfo{author}{\bibfnamefont{B.~E.} \bibnamefont{Feldman}},
  \bibinfo{author}{\bibfnamefont{R.~T.} \bibnamefont{Weitz}},
  \bibinfo{author}{\bibfnamefont{M.~T.} \bibnamefont{Allen}}, \bibnamefont{and}
  \bibinfo{author}{\bibfnamefont{A.}~\bibnamefont{Yacoby}},
  \bibinfo{journal}{Phys. Rev. Lett.} \textbf{\bibinfo{volume}{105}},
  \bibinfo{pages}{256806} (\bibinfo{year}{2010}).

\bibitem[{\citenamefont{Weitz et~al.}(2010)\citenamefont{Weitz, Allen, Feldman,
  Martin, and Yacoby}}]{weitz2010}
\bibinfo{author}{\bibfnamefont{R.~T.} \bibnamefont{Weitz}},
  \bibinfo{author}{\bibfnamefont{M.~T.} \bibnamefont{Allen}},
  \bibinfo{author}{\bibfnamefont{B.~E.} \bibnamefont{Feldman}},
  \bibinfo{author}{\bibfnamefont{J.}~\bibnamefont{Martin}}, \bibnamefont{and}
  \bibinfo{author}{\bibfnamefont{A.}~\bibnamefont{Yacoby}},
  \bibinfo{journal}{Science} \textbf{\bibinfo{volume}{330}},
  \bibinfo{pages}{812} (\bibinfo{year}{2010}).

\bibitem[{\citenamefont{{Mayorov} et~al.}(2011)\citenamefont{{Mayorov},
  {Elias}, {Mucha-Kruczynski}, {Gorbachev}, {Tudorovskiy}, {Zhukov}, {Morozov},
  {Katsnelson}, {Geim}, and {Novoselov}}}]{mayorov2011}
\bibinfo{author}{\bibfnamefont{A.~S.} \bibnamefont{{Mayorov}}},
  \bibinfo{author}{\bibfnamefont{D.~C.} \bibnamefont{{Elias}}},
  \bibinfo{author}{\bibfnamefont{M.}~\bibnamefont{{Mucha-Kruczynski}}},
  \bibinfo{author}{\bibfnamefont{R.~V.} \bibnamefont{{Gorbachev}}},
  \bibinfo{author}{\bibfnamefont{T.}~\bibnamefont{{Tudorovskiy}}},
  \bibinfo{author}{\bibfnamefont{A.}~\bibnamefont{{Zhukov}}},
  \bibinfo{author}{\bibfnamefont{S.~V.} \bibnamefont{{Morozov}}},
  \bibinfo{author}{\bibfnamefont{M.~I.} \bibnamefont{{Katsnelson}}},
  \bibinfo{author}{\bibfnamefont{A.~K.} \bibnamefont{{Geim}}},
  \bibnamefont{and} \bibinfo{author}{\bibfnamefont{K.~S.}
  \bibnamefont{{Novoselov}}}, \bibinfo{journal}{Science}
  \textbf{\bibinfo{volume}{333}}, \bibinfo{pages}{860} (\bibinfo{year}{2011}).

\bibitem[{\citenamefont{{Freitag} et~al.}(2012)\citenamefont{{Freitag},
  {Trbovic}, {Weiss}, and {Sch{\"o}nenberger}}}]{freitag2011}
\bibinfo{author}{\bibfnamefont{F.}~\bibnamefont{{Freitag}}},
  \bibinfo{author}{\bibfnamefont{J.}~\bibnamefont{{Trbovic}}},
  \bibinfo{author}{\bibfnamefont{M.}~\bibnamefont{{Weiss}}}, \bibnamefont{and}
  \bibinfo{author}{\bibfnamefont{C.}~\bibnamefont{{Sch{\"o}nenberger}}},
  \bibinfo{journal}{Phys. Rev. Lett.} \textbf{\bibinfo{volume}{108}},
  \bibinfo{eid}{076602} (\bibinfo{year}{2012}).

\bibitem[{\citenamefont{{Velasco} et~al.}(2012)\citenamefont{{Velasco}, {Jing},
  {Bao}, {Lee}, {Kratz}, {Aji}, {Bockrath}, {Lau}, {Varma}, {Stillwell}
  et~al.}}]{velasco2011}
\bibinfo{author}{\bibfnamefont{J.~J.} \bibnamefont{{Velasco}}},
  \bibinfo{author}{\bibfnamefont{L.}~\bibnamefont{{Jing}}},
  \bibinfo{author}{\bibfnamefont{W.}~\bibnamefont{{Bao}}},
  \bibinfo{author}{\bibfnamefont{Y.}~\bibnamefont{{Lee}}},
  \bibinfo{author}{\bibfnamefont{P.}~\bibnamefont{{Kratz}}},
  \bibinfo{author}{\bibfnamefont{V.}~\bibnamefont{{Aji}}},
  \bibinfo{author}{\bibfnamefont{M.}~\bibnamefont{{Bockrath}}},
  \bibinfo{author}{\bibfnamefont{C.~N.} \bibnamefont{{Lau}}},
  \bibinfo{author}{\bibfnamefont{C.}~\bibnamefont{{Varma}}},
  \bibinfo{author}{\bibfnamefont{R.}~\bibnamefont{{Stillwell}}},
  \bibnamefont{et~al.}, \bibinfo{journal}{Nature Nanotechnology}
  \textbf{\bibinfo{volume}{7}}, \bibinfo{pages}{156} (\bibinfo{year}{2012}).

\bibitem[{\citenamefont{Bao et~al.}(2012)\citenamefont{Bao, Velasco~Jr., Zhang,
  Jing, Standley, Smirnov, Bockrath, MacDonald, and Lau}}]{bao2012}
\bibinfo{author}{\bibfnamefont{W.}~\bibnamefont{Bao}},
  \bibinfo{author}{\bibfnamefont{J.}~\bibnamefont{Velasco~Jr.}},
  \bibinfo{author}{\bibfnamefont{F.}~\bibnamefont{Zhang}},
  \bibinfo{author}{\bibfnamefont{L.}~\bibnamefont{Jing}},
  \bibinfo{author}{\bibfnamefont{B.}~\bibnamefont{Standley}},
  \bibinfo{author}{\bibfnamefont{D.}~\bibnamefont{Smirnov}},
  \bibinfo{author}{\bibfnamefont{M.}~\bibnamefont{Bockrath}},
  \bibinfo{author}{\bibfnamefont{A.~H.} \bibnamefont{MacDonald}},
  \bibnamefont{and} \bibinfo{author}{\bibfnamefont{C.~N.} \bibnamefont{Lau}},
  \bibinfo{journal}{Proc. Natl. Acad. Sci. USA} \textbf{\bibinfo{volume}{109}},
  \bibinfo{pages}{10802} (\bibinfo{year}{2012}).

\bibitem[{\citenamefont{Veligura et~al.}(2012)\citenamefont{Veligura, van
  Elferen, Tombros, Maan, Zeitler, and van Wees}}]{veligura2012}
\bibinfo{author}{\bibfnamefont{A.}~\bibnamefont{Veligura}},
  \bibinfo{author}{\bibfnamefont{H.~J.} \bibnamefont{van Elferen}},
  \bibinfo{author}{\bibfnamefont{N.}~\bibnamefont{Tombros}},
  \bibinfo{author}{\bibfnamefont{J.~C.} \bibnamefont{Maan}},
  \bibinfo{author}{\bibfnamefont{U.}~\bibnamefont{Zeitler}}, \bibnamefont{and}
  \bibinfo{author}{\bibfnamefont{B.~J.} \bibnamefont{van Wees}},
  \bibinfo{journal}{Phys. Rev. B} \textbf{\bibinfo{volume}{85}},
  \bibinfo{pages}{155412} (\bibinfo{year}{2012}).

\bibitem[{\citenamefont{Freitag et~al.}(2013)\citenamefont{Freitag, Weiss,
  Maurand, Trbovic, and Sch{\"o}nenberger}}]{freitag2013}
\bibinfo{author}{\bibfnamefont{F.}~\bibnamefont{Freitag}},
  \bibinfo{author}{\bibfnamefont{M.}~\bibnamefont{Weiss}},
  \bibinfo{author}{\bibfnamefont{R.}~\bibnamefont{Maurand}},
  \bibinfo{author}{\bibfnamefont{J.}~\bibnamefont{Trbovic}}, \bibnamefont{and}
  \bibinfo{author}{\bibnamefont{Sch{\"o}nenberger}}, \bibinfo{journal}{Phys.
  Rev. B} \textbf{\bibinfo{volume}{87}}, \bibinfo{pages}{161402(R)}
  (\bibinfo{year}{2013}).

\bibitem[{\citenamefont{Cvetkovic et~al.}(2012)\citenamefont{Cvetkovic,
  Throckmorton, and Vafek}}]{cvetkovic2012}
\bibinfo{author}{\bibfnamefont{V.}~\bibnamefont{Cvetkovic}},
  \bibinfo{author}{\bibfnamefont{R.~E.} \bibnamefont{Throckmorton}},
  \bibnamefont{and} \bibinfo{author}{\bibfnamefont{O.}~\bibnamefont{Vafek}},
  \bibinfo{journal}{Phys. Rev. B} \textbf{\bibinfo{volume}{86}},
  \bibinfo{pages}{075467} (\bibinfo{year}{2012}).

\bibitem[{\citenamefont{Abrikosov and Gorkov}(1961)}]{abrikosov1961}
\bibinfo{author}{\bibfnamefont{A.~A.} \bibnamefont{Abrikosov}}
  \bibnamefont{and} \bibinfo{author}{\bibfnamefont{L.~P.}
  \bibnamefont{Gorkov}}, \bibinfo{journal}{Soviet Phys. JETP}
  \textbf{\bibinfo{volume}{12}}, \bibinfo{pages}{1243} (\bibinfo{year}{1961}).

\bibitem[{\citenamefont{Maki}(1969)}]{maki1969}
\bibinfo{author}{\bibfnamefont{K.}~\bibnamefont{Maki}}, in
  \emph{\bibinfo{booktitle}{Superconductivity}}, edited by
  \bibinfo{editor}{\bibfnamefont{R.~D.} \bibnamefont{Parks}}
  (\bibinfo{publisher}{Dekker, New York}, \bibinfo{year}{1969}).

\bibitem[{\citenamefont{Zittartz}(1967)}]{zittartz1967}
\bibinfo{author}{\bibfnamefont{J.}~\bibnamefont{Zittartz}},
  \bibinfo{journal}{Phys. Rev.} \textbf{\bibinfo{volume}{164}},
  \bibinfo{pages}{575} (\bibinfo{year}{1967}).

\bibitem[{\citenamefont{{Bistritzer} and {MacDonald}}(2008)}]{bistritzer2008a}
\bibinfo{author}{\bibfnamefont{R.}~\bibnamefont{{Bistritzer}}}
  \bibnamefont{and} \bibinfo{author}{\bibfnamefont{A.~H.}
  \bibnamefont{{MacDonald}}}, \bibinfo{journal}{Phys. Rev. Lett.}
  \textbf{\bibinfo{volume}{101}}, \bibinfo{pages}{256406}
  (\bibinfo{year}{2008}).

\end{thebibliography}
\end{document}